\documentclass[11pt,a4paper]{article}
\usepackage{url}
\usepackage{breakurl}
\usepackage[hyperref]{ranlp2021}
\usepackage{times}
\usepackage{latexsym}
\usepackage{amsmath}
\usepackage{adjustbox}

\usepackage{microtype}

\aclfinalcopy 

\newcommand{\specialcell}[2][c]{%
  \begin{tabular}[#1]{@{}c@{}}#2\end{tabular}}

\title{Semi-Automated Labeling of Requirement Datasets for Relation Extraction}
\author{Jeremias Bohn \\
  Technical University of Munich, Germany \\
  \texttt{jeremias.bohn@tum.de} \\\And
  Jannik Fischbach \\
  Qualicen GmbH, Germany \\
  \texttt{jannik.fischbach@qualicen.de} \\\AND
  Martin Schmitt \\
  Center for Information and Language Processing \\
  LMU Munich, Germany \\
  \texttt{martin@cis.lmu.de}\\\AND
  Hinrich Schütze \\
  Center for Information and Language Processing \\
  LMU Munich, Germany \\
  \texttt{inquiries@cislmu.org}\\\And
  Andreas Vogelsang \\
  University of Cologne, Germany \\
  \texttt{vogelsang@cis.uni-koeln.de} \\}

\date{July 5, 2021}

\begin{document}
\maketitle
\begin{abstract}
Creating datasets manually by human annotators is a laborious task that can lead to biased and inhomogeneous labels. We propose a flexible, semi-automatic framework for labeling data for relation extraction. Furthermore, we provide a dataset of preprocessed sentences from the requirements engineering domain, including a set of automatically created as well as hand-crafted labels. In our case study, we compare the human and automatic labels and show that there is a substantial overlap between both annotations.
\end{abstract}
\section{Introduction}
While recent advances in Natural Language Processing have yielded high-quality language models such as BERT~\cite{devlin-etal-2019-bert}, GPT-3~\cite{brown2020language} and ELECTRA \cite{clark2020electra} which are able to continue sentences, fill in masked words and correctly parse human language, using these models for most use-case scenarios still requires them to be trained on a down-stream task using labeled data. For some tasks, e.g. sentiment analysis of reviews, creating datasets is relatively easy as large databases with annotations already exist (such as the IMDb movie review dataset~\cite{maas-EtAl:2011:ACL-HLT2011}). However, training a model on niche tasks often demands hand-crafting new datasets from spread-out documents. This is usually done by humans who collect, preprocess, and annotate sentences which is a laborious task and can result in biased and/or inhomogeneous labeling, e.g. if annotation instructions were not understood correctly or left room for subjective interpretation. This becomes especially apparent if multiple, non-expert individuals are involved in this process.\\
In requirements engineering, we usually work with large documents written in natural language~\cite{Mich04,Kassab14} which describe the specifications of a software project, usually classified as either functional requirements, specifying what functionality the system should provide, and non-functional requirements, specifying in what way the system should implement those functions. However, these documents are often updated during the life cycle of the project and span up to multiple hundreds of pages, depending on the project size. Keeping track of all the changes and maintaining the software based on the requirement document can soon become a challenge~\cite{fischbach20} which is why an automatic conversion to, e.g., UML diagrams can come in handy. To do so, it is necessary to parse the relations between entities from the written text into a structured format, thus creating a comparable corpus of requirements in natural language and the same relation in a formal language.\\
In this paper, we propose a semi-automatic approach that, given a clean, grammatically correct sentence stating a software requirement, outputs a labeling corresponding to the relation the requirement describes based on a small set of pre-defined rules of word dependency relations. This should reduce human bias manifesting in labels as the annotator does not actively choose the labels for each word anymore but instead defines abstract rules which provide for homogeneous, deterministic labeling and reduce the amount of labor for creating such datasets. This automatically annotated data can then be used for training a more powerful model, as shown by~\citet{schmitt-etal-2020-unsupervised}.\\
We summarize our main contributions as follows:
\begin{itemize}
    \item We provide a high-quality, preprocessed dataset of 2,093 requirement sentences together with 1,848 automatically created labels and another 199 manually created labels for a subset of the automatically labeled sentences as a resource for further research projects.
    \item We provide a flexible, semi-automatic framework for data annotation of the relation extraction domain based on dependency parsing and pattern matching.
    \item We conduct a case study on the said framework on requirement document sentences, showing its annotation results are matching those of humans to a substantial degree.
\end{itemize}
\section{Related Work}
\citet{gamallo-etal-2012-dependency} propose a simple Open Information Extraction system based on dependency parse trees. The algorithm extracts triples with two arguments and a sentence part relating those. However, the patterns are not very sophisticated and put a large part of the sentence into the relation. Hence, this approach is not suitable for our use case as we would eventually like to generate object diagrams from the relations we extracted.
\citet{erkan-etal-2007-semi} use dependency parse trees to extract relations between proteins from sentences. They do so by classifying whether a sentence, given a dependency tree, describes a relation between any pair of proteins occurring in the sentence using semi-supervised harmonic functions and support vector machines. However, their entities (the protein names) are already annotated which is not the case if we only have the raw sentences as in our approach.
\citet{mausam-etal-2012-open} use dependency trees and a labeled bootstrap dataset to automatically generate patterns for information extraction, unlike our approach which does not require annotating any data manually but instead to produce patterns. While this approach might be able to extract simple triples well, one needs either a larger annotated dataset, defeating the purpose of our work, or the patterns might not generalize well, thus being unsuitable for constructing a qualitative annotated corpus.
\citet{reddy-etal-2016-transforming} propose an algorithm to automatically extract logical expressions from dependency parse trees for question answering. These were then converted into a graph indicating the relations between the named entities in the sentence by applying semantic parsing. However, this approach always converts the entire sentence into a graph and may include information that is irrelevant for a dataset that is to be generated.
\citet{inago2019parsing} use a rule-based approach on dependency trees to process natural language car parking instructions with decision trees for automated driving systems. Unlike our data (or most datasets in general), sentences of the application domain are very short and similar in structure. While our approach could be effectively converted into a decision tree, it is easier to construct rules with our pattern engine for more complex data.
\section{Corpus Creation}

\subsection{Dataset} For our dataset, we use 19 publicly available requirement documents in the English language from the PURE dataset~\cite{ferrari2017pure}, with a large topical variety, including governmental institution software in military and scientific fields, inventory management systems and video games. All documents are provided in .PDF, .HTML or .DOC format. From these, we manually extracted 2,104 requirement sentences (1,639 functional, 465 non-functional requirements).
\subsection{Preprocessing} As we want to automatically dependency parse our sentences, we have to ensure that all input to the model is grammatically and orthographically sound. We also have to ensure that any unnecessary information is removed to not confuse the parser. Therefore, we manually applied the following formatting operations to each sentence during data extraction:
\begin{itemize}
    \item Splitting of enumerations into multiple sentences, adjusting words if necessary to make the sentence sound (e.g., nounification of verbs); e.g., "The system has to include a) [...] b) [...] c) [...]" becomes 3 sentences, each including exactly one of the requirements
    \item Removal of extra inter-punctuation (additional spaces, dots, commas, etc.)
    \item Removal of references to sections, tables, figures, or other requirements of the document as they are not relevant for extracting the relation of the sentence itself
    \item Removal of abbreviations after written-out expressions (e.g., in "automated teller machine (ATM)", the "(ATM)" is dropped)
    \item Removal of requirement reference numbers
    \item Correction of spelling mistakes where obvious
    \item Adding of dots at the end of each sentence if missing
    \item Changing the first letter of a sentence to upper case if it is not yet
    \item Removal of quotation marks around pseudo-correct terms (e.g., 'the "processor" will [...]' becomes 'the processor will [...]')
    \item Removal of explicit explanations of what is included in some term (e.g., "errors of either kind, i.e. hardware and software, [...]")
    \item Lower-casing of words if they are not abbreviations (e.g., "NOT" becomes "not")
    \item Remove brackets around additional plural 's' (e.g., "socket(s)" becomes "sockets")
    \item Exchanging "/" with "and" or "or" where applicable and possible given the context (e.g. "The system should support adding/deleting files" becomes "The system should support adding and deleting files")
    \item Unification of the possessive 's' preceding symbols ("`" and "´" are changed to "'")
    \item Removal of duplicate sentences (11 in total)
\end{itemize}
After these preprocessing steps, the average sentence length is 19.87 words, the maximum is 69 words and the minimum 4 words.
\subsection{Labeling} These final 2,093 sentences (1,628 functional, 465 non-functional requirements) are parsed to extract dependencies using the Neural Adobe-UCSD Parser~\cite{mrini-etal-2020-rethinking} which achieved state-of-the-art performance on the Penn Treebank dataset \cite{marcus-etal-1993-building}. Based on these dependencies, we handcraft a total of 102 patterns to label 91.03\% of the functional and 78.71\% of the non-functional sentences without any further human interaction. Each pattern is a sequence of triples $(l, dp, c)$ where $l$ is a label, $dp$ a sequence of dependency labels forming a path downwards a dependency tree and $c$ a Boolean value indicating whether all children (direct and indirect) should be left out from labeling or not. Each sequence applies all or a subset of the following entity tags to the sentences:
\begin{itemize}
    \item \texttt{ent1}: The main entity of the requirement. Either the acting component or the component on which a constraint is applied (if there is no second entity)
    \item \texttt{rel}: The relation/action of the requirement.
    \item \texttt{ent2}: The passive entity of the requirement. Either the component on which an action is performed or which is involved in the action passively
    \item \texttt{cond}: Any modifier of the requirement. Can further specify the requirement or put conditions on it how or when it will be applied.
\end{itemize}
An excerpt of automatic annotations can be found in Table~\ref{tab:labeling}.
\begin{table*}[htpb]
    \centering
    \begin{adjustbox}{width=\textwidth}
    \begin{tabular}{| c |}
        \hline
         \textbf{Sentence} \\
        \hline
        \\
          $\text{\underline{While flying two MAE AVs Beyond Line Of Sight} }_\texttt{cond}\text{, \underline{the TCS} }_{\texttt{ent1}} \text{ shall \underline{provide} }_{\texttt{rel}} \text{ \underline{full control functionality} }_\texttt{ent2}\text{ \underline{of each AV} }_{\texttt{cond}}\text{.}$\\
         $\text{\underline{NPAC SMS} }_\texttt{ent1}\text{ shall \underline{default} }_\texttt{rel}\text{ \underline{the EDR Indicator} }_\texttt{ent2}\text{ \underline{to False} }_\texttt{cond} \text{.}$ \\
         $\text{\underline{A bulk entry} }_\texttt{ent1}\text{ can be used to \underline{add} }_\texttt{rel}\text{ \underline{many assets} }_\texttt{ent2}\text{.}$\\
         $\text{\underline{The HATS-GUI} }_\texttt{ent1}\text{ shall interact with the Host OS to \underline{compare} }_\texttt{rel}\text{ \underline{time stamps} }_\texttt{ent2}\text{ \underline{ for files} }_\texttt{cond}\text{.}$\\
         $\text{\underline{The BE} }_\texttt{ent1}\text{ shall be able to \underline{apply} }_\texttt{rel}\text{ \underline{corrections} }_\texttt{ent2}\text{ \underline{based on state count and/or quantizer power measurement data} }_\texttt{cond}\text{.}$\\\\
         \hline
    \end{tabular}
    \end{adjustbox}
    \caption{\label{tab:labeling} Examples of Labeling}
    
\end{table*}
Each pattern is applied using tree traversal: for each label that is to be applied, a sequence of dependency labels (optionally with modifiers) is given, starting at the root. The algorithm checks whether the current nodes have any direct children connected to them with the current dependency label of the sequence. If so, we check whether these children have children connected to them with the next label in the sequence. If not, the pattern fitting is stopped and no labeling is applied to the sentence. If we reach the end of the sequence, the final node is labeled with the given label and, depending on a parameter, all of its children, too. A simple example can be found in Table \ref{tab:patterns}, row 1.
Dependency labels can include modifiers to allow for more complex patterns:
\begin{itemize}
    \item Starting with \texttt{!}, the pattern matching will remove any node that has one or more children with the given dependency label. Thus, no step downwards the tree is taken
    \item Followed by \texttt{=[placeholder]} where \texttt{[placeholder]} is any word, only those nodes are considered where the label is the given label and the actual word of the node is specified by \texttt{[placeholder]}
    \item \texttt{..} lets us traverse back to the parent of the current node. This allows us to check nodes for their existence without including them in the actual labeling 
\end{itemize}
A selection of patterns used can be found in Table~\ref{tab:patterns}. In our setting, one sentence usually holds one relation, however, this is not the case for conjunctions of multiple main clauses or instructions. Due to current limitations of our engine (see Section \ref{section:conclusion_outlook}), the relation of the first main clause is always chosen, however, this depends on the pattern design. Even though we only use requirements written in English, a large portion of the rules could be applied to data in different languages as the Universal Dependencies~\cite{schuster-manning-2016-enhanced} rely on the concept of primacy of content, allowing for very similar dependency trees. However, patterns explicitly using keywords may not generalize well for other languages. The code for the labeling task as well as the labeled data can be found on GitHub\footnote{\url{https://github.com/JeremiasBohn/RequirementRelationExtractor}}.
\begin{table*}[htpb]
    \centering
    \begin{adjustbox}{width=\textwidth}
    \begin{tabular}{| c | c |}
        \hline
         \textbf{Pattern} & \textbf{Description} \\
        \hline
        \specialcell{
         ('rel', ['root'], True)\\
        ('ent1', ['root', 'nsubj'], False)\\
        ('ent2', ['root', 'dobj'], False)\\
        ('cond', ['root', 'advcl'], False)} & \specialcell{Simple pattern, sets the root of the sentence as\\
        the relation (only this single word), the entire nominal subject\\as the acting entity, the entire direct object as\\ the passive entity. An adverbial clause is treated as a\\ relation modifier.} \\
        \hline
         \specialcell{('rel', ['root=capable', 'prep=of', 'pcomp'], True)\\
        ('ent1', ['root', 'nsubj'], False)\\
        ('ent2', ['root', 'prep=of', 'pcomp', 'prep=in', 'pobj'], False)\\
        ('cond', ['root', 'advcl'], False)} & \specialcell{Catches phrases like "The system should be capable of [...]"\\and searches for the passive entity in the prepositional object of\\the prepositional clause starting with "in".} \\
        \hline
        \specialcell{('rel', ['root', '!dobj'], True)\\
        ('ent1', ['root', 'nsubjpass'], False)\\
        ('cond', ['root', 'prep=in', 'pobj=case', '..'], False)\\
        ('cond', ['root', 'advmod'], False)} & \specialcell{Pattern is only applied if the sentence has\\no direct object (which could serve as the passive entity).\\Prepositional sentences starting with "in case" are\\labeled as requirement modifier (we have to traverse\\the tree upwards again to include the 'in' as well).}\\
         \hline
    \end{tabular}
    \end{adjustbox}
    \caption{Examples of Patterns}
    \label{tab:patterns}
\end{table*}
\section{Evaluation}
Given our automatically labeled data, we evaluate the quality of the labels by comparing its output to human annotations. To do so, we randomly sample 199 sentences (10.77\%) from the 1848 sentences which were automatically labeled. Two of the authors then annotated these sentences manually. The annotators were given the descriptions of each label type, but had no access to the actual labeling from the algorithm. Annotators collaboratively labeled the data, discussing the labeling for each sentence and agreeing upon a single valid labeling. We then calculate inter-rater reliability with the Cohen's $\kappa$ between the human annotators and the automatic annotator, once over all labels and once as average inter-reliability per sentence (i.e., we calculate one Cohen's $\kappa$ score per sentence and average over all sentences \textendash this considers each sentence equally while the overall score puts more weight on longer sentences). The results can be found in Table~\ref{tab:eval}.
\begin{table}[htpb]
    \centering
    \begin{tabular}{| c | c | c |}
    \hline
         \textbf{Labels considered} & \textbf{Sentence Avg.}  & \textbf{Overall} \\
         \hline
         All labels &0.632& 0.576 \\
         \texttt{rel} only &0.790& 0.720\\
         \texttt{ent1} only &0.855& 0.822\\
         \texttt{ent2} only &0.619& 0.561\\
         \texttt{cond} only &0.532& 0.543\\
         \hline
    \end{tabular}
    \caption{Cohen's Kappa Results}
    \label{tab:eval}
\end{table}
While the overall score puts more weight on long sentences, the sentence average provides us an approximation of the reliability of our automatic annotator for any sentence. According to the taxonomy of Landis and Koch~\cite{landis1977measurement}, the per sentence average $\kappa$ value indicates a substantial inter-annotator agreement, the overall $\kappa$ a moderate agreement. While the main acting entity is extracted very well with almost perfect agreement according to Landis and Koch, extracting relational modifiers proofs to be the hardest with only moderate agreement between our automatic approach and the human annotators. This is mostly due to the nature of the label itself, spanning a large variety of modifiers from conditions to entities not involved in the relation itself. While one could split the \texttt{cond} label into multiple different labels, this would greatly increase the number of patterns required. Alternatively, one might reduce the coverage of the labeling in general but we focused on including as much information as possible. The relatively low score for \texttt{ent2} mainly arises from sentences containing multiple relations where many words describe a passive entity for other relations than the one of the main sentence. Our approach currently is not able to effectively extract multiple relations from a single sentence yet. This is also the reason why the score \texttt{rel} is lower than the one for \texttt{ent1}.
\section{Limitations}
While our approach works well for requirements documents - after all, relations between software entities and modifications of these relations can be extracted well by syntactically parsing the sentence structure - this does not apply to word labels which require a semantic understanding of the input. For example, if we were to create labels for Named Entity Recognition, our algorithm would fail as it is not possible to find syntactic rules to distinguish between, e.g., an organization and a person. Also, the algorithm fails in some cases if either rules are not specific enough or the dependency parser mistakenly adds dependencies between sentence parts where there is no dependency between them. The latter may especially occur frequently if the sentences were not preprocessed well which is why our algorithm is not suitable as a classifier in general (if we, on the other hand, use our data as training input for a Transformer model~\cite{vaswani2017attention}, it may overcome these strict syntactic requirements and generalize better on real-world data).
\section{Conclusion \& Outlook}\label{section:conclusion_outlook}
In this paper, we present a novel approach for data labeling which allows users to annotate sentences for relation extraction within a shorter time period compared to manual annotation while at the same time having a consistent labeling scheme for the entire dataset. Our approach exploits syntactic features which are the integral foundation of most relation extraction tasks.\\
For the future, it would be helpful to implement an automatic extraction of requirement sentences by, e.g., training a classifier to identify relevant sentences in plain text or .PDF documents as well as a semi-automatic approach with human validation for preprocessing sentences into grammatically and orthographically sound ones. We plan on extending the pattern engine our algorithm relies on, e.g., allowing for recursive patterns to parse nested sentences and to extract multiple relations from one sentence as well as optional pattern parts to reduce redundancy (e.g., a sentence where the active entity is the nominal subject, the relation the dependency tree root and the passive entity the direct object may have a relation modifier in an adverbial clause. As of the current state, this requires two patterns (exponentially increasing with the number of optional dependencies) while with a pattern where this adverbial clause is considered optional, we only need a single pattern).

\bibliographystyle{acl_natbib}
\bibliography{anthology,ranlp2021}


\end{document}